\documentstyle[12pt]{article}

\skewchar\fivmi='177
\skewchar\sixmi='177
\skewchar\sevmi='177
\skewchar\egtmi='177
\skewchar\ninmi='177
\skewchar\tenmi='177
\skewchar\elvmi='177
\skewchar\twlmi='177
\skewchar\frtnmi='177
\skewchar\svtnmi='177
\skewchar\twtymi='177
\def\@magscale#1{ scaled \magstep #1}
\skewchar\fivsy='60
\skewchar\sixsy='60
\skewchar\sevsy='60
\skewchar\egtsy='60
\skewchar\ninsy='60
\skewchar\tensy='60
\skewchar\elvsy='60
\skewchar\twlsy='60
\skewchar\frtnsy='60
\skewchar\svtnsy='60
\skewchar\twtysy='60

\catcode`@=11
\def\un#1{\relax\ifmmode\@@underline#1\else
        $\@@underline{\hbox{#1}}$\relax\fi}
\catcode`@=12

\def\dslash{\not{\hbox{\kern-2pt $\partial$}}}
\def\Dslash{\not{\hbox{\kern-4pt $D$}}}
\def\pslash{\not{\hbox{\kern-2.3pt $p$}}}
 \newtoks\slashfraction
 \slashfraction={.13}
 \def\slash#1{\setbox0\hbox{$ #1 $}
 \setbox0\hbox to \the\slashfraction\wd0{\hss \box0}/\box0 }

\font\ro=cmsy10                          
\def\kcr{{\hbox{\ro \char'170}}}                
\def\ktl{{\hbox{\ro \char'170}}}        
\def\ktr{{\hbox{\ro \char'170}}}        
\def\kbl{{\hbox{\ro \char'170}}}        
\def\kbr{{\hbox{\ro \char'170}}}        

\def\plpl{\raise-2pt\hbox{$\raise3pt\hbox{$_+$}\hskip-6.67pt\raise0.0pt
\hbox{$^+$}\hskip 0.01pt$}}
\def\mimi{\raise-2pt\hbox{$\raise3pt\hbox{$_-$}\hskip-6.67pt\raise0.0pt
\hbox{$^-$}\hskip 0.01pt$}}

\def\bo{{\raise.15ex\hbox{\large$\Box$}}}               
\def\TH{{\raise.2ex\hbox{$\displaystyle \bigodot$}\mskip-4.7mu \llap H \;}}
\def\face{{\raise.2ex\hbox{$\displaystyle \bigodot$}\mskip-2.2mu \llap {$\ddot
        \smile$}}}                                      

\def\sp#1{{}^{#1}}                              
\def\leftrightarrowfill{$\mathsurround=0pt \mathord\leftarrow \mkern-6mu
        \cleaders\hbox{$\mkern-2mu \mathord- \mkern-2mu$}\hfill
        \mkern-6mu \mathord\rightarrow$}
\def\dvec#1{\vbox{\ialign{##\crcr
        \leftrightarrowfill\crcr\noalign{\kern-1pt\nointerlineskip}
        $\hfil\displaystyle{#1}\hfil$\crcr}}}           

\def\frac#1#2{{\textstyle{#1\over\vphantom2\smash{\raise.20ex
        \hbox{$\scriptstyle{#2}$}}}}}                   
\def\sfrac#1#2{{\vphantom1\smash{\lower.5ex\hbox{\small$#1$}}\over
        \vphantom1\smash{\raise.4ex\hbox{\small$#2$}}}} 
\def\bfrac#1#2{{\vphantom1\smash{\lower.5ex\hbox{$#1$}}\over
        \vphantom1\smash{\raise.3ex\hbox{$#2$}}}}       
\def\afrac#1#2{{\vphantom1\smash{\lower.5ex\hbox{$#1$}}\over#2}}    

\newskip\humongous \humongous=0pt plus 1000pt minus 1000pt

\newif\ifdtup

\def\ref#1{$\sp{#1)}$}

\parskip=\medskipamount                 
\thicklines                         

\def\oldheadpic{                                
        \setlength{\unitlength}{.4mm}
        \thinlines
        \par
        \begin{picture}(349,16)
        \put(325,16){\line(1,0){4}}
        \put(330,16){\line(1,0){4}}
        \put(340,16){\line(1,0){4}}
        \put(335,0){\line(1,0){4}}
        \put(340,0){\line(1,0){4}}
        \put(345,0){\line(1,0){4}}
        \put(329,0){\line(0,1){16}}
        \put(330,0){\line(0,1){16}}
        \put(339,0){\line(0,1){16}}
        \put(340,0){\line(0,1){16}}
        \put(344,0){\line(0,1){16}}
        \put(345,0){\line(0,1){16}}
        \put(329,16){\oval(8,32)[bl]}
        \put(330,16){\oval(8,32)[br]}
        \put(339,0){\oval(8,32)[tl]}
        \put(345,0){\oval(8,32)[tr]}
        \end{picture}
        \par
        \thicklines
        \vskip.2in}
\def\oldtitle#1#2#3#4{\oldheadpic\begin{center}\vglue.5in{\large\bf #1}\\[.6in]
        {#2}\\[.1in] {\it Department of Physics and Astronomy}\\
        {\it University of Maryland, College Park, MD 20742}\\[.6in]
        Physics Publication \#{#3}\\ {#4}\\[1.5in] {\bf ABSTRACT}\\[.1in]
        \end{center} \begin{quotation}}                 
\def\oldTitle#1#2#3#4#5#6#7{\oldheadpic\begin{center} \vglue .4in
        {\large\bf #1}\\[.4in]
        {#2}\\[.1in] {\it Department of Physics and Astronomy}\\
        {\it University of Maryland, College Park, MD 20742}\\[.1in]
        {#3}\\[.1in] {\it {#4}}\\ {\it {#5}}\\[.4in]
        Physics Publication \#{#6}\\ {#7}\\[.5in] {\bf ABSTRACT}\\[.1in]
        \end{center} \begin{quotation}}                 
\def\border{                                            
        \setlength{\unitlength}{1mm}
        \newcount\xco
        \newcount\yco
        \xco=-21
        \yco=12
        \begin{picture}(140,0)
        \put(\xco,\yco){$\ktl$}
        \advance\yco by-1
        {\loop
        \put(\xco,\yco){$\kcr$}
        \advance\yco by-2
        \ifnum\yco>-240
        \repeat
        \put(\xco,\yco){$\kbl$}}
        \xco=158
        \yco=12
        \put(\xco,\yco){$\ktr$}
        \advance\yco by-1
        {\loop
        \put(\xco,\yco){$\kcr$}
        \advance\yco by-2
        \ifnum\yco>-240
        \repeat
        \put(\xco,\yco){$\kbr$}}
        \put(-20,13){\tiny University of Maryland Elementary Particle
Physics University of Maryland Elementary Particle Physics University of
Maryland Elementary Particle Physics}
        \put(-20,-241.5){\tiny University of Maryland Elementary
Particle Physics University of Maryland Elementary Particle Physics
University of Maryland Elementary Particle Physics}
        \end{picture}
        \par\vskip-8mm}
\def\bordero{                                           
        \setlength{\unitlength}{1mm}
        \newcount\xco
        \newcount\yco
        \xco=-31
        \yco=12
        \begin{picture}(140,0)
        \put(\xco,\yco){$\ktl$}
        \advance\yco by-1
        {\loop
        \put(\xco,\yco){$\kclr}
        \advance\yco by-2
        \ifnum\yco>-240
        \repeat
        \put(\xco,\yco){$\kbl$}}
        \xco=151
        \yco=12
        \put(\xco,\yco){$\ktr$}
        \advance\yco by-1
        {\loop
        \put(\xco,\yco){$\kcr$}
        \advance\yco by-2
        \ifnum\yco>-240
        \repeat
        \put(\xco,\yco){$\kbr$}}
        \put(-20,12){\ooo
bacdefghidfghghdhededbihdgdfdfhhdheidhdhebaaahjhhdahba

hgdedge
   hgfdiehhgdigicba}
        \put(-20,-241.5){\ooo
ababaighefdbfghgeahgdfgafagihdidihiidhiagfedhadbfd

ecdcdfa
   gdcbhaddhbgfchbgfdacfediacbabab}
        \end{picture}
        \par\vskip-8mm}
\def\headpic{                                           
        \indent
        \setlength{\unitlength}{.4mm}
        \thinlines
        \par
        \begin{picture}(29,16)
        \put(165,16){\line(1,0){4}}
        \put(170,16){\line(1,0){4}}
        \put(180,16){\line(1,0){4}}
        \put(175,0){\line(1,0){4}}
        \put(180,0){\line(1,0){4}}
        \put(185,0){\line(1,0){4}}
        \put(169,0){\line(0,1){16}}
        \put(170,0){\line(0,1){16}}
        \put(179,0){\line(0,1){16}}
        \put(180,0){\line(0,1){16}}
        \put(184,0){\line(0,1){16}}
        \put(185,0){\line(0,1){16}}
        \put(169,16){\oval(8,32)[bl]}
        \put(170,16){\oval(8,32)[br]}
        \put(179,0){\oval(8,32)[tl]}
        \put(185,0){\oval(8,32)[tr]}
        \end{picture}
        \par\vskip-6.5mm
        \thicklines}
\def\title#1#2#3#4{\border\headpic {\hbox to\hsize{#4 \hfill UMDEPP #3}}\par
        \begin{center} \vglue .5in {\large\bf #1}\\[.6in]
        {#2}\\[.1in] {\it Department of Physics and Astronomy}\\
        {\it University of Maryland, College Park, MD 20742}\\[1.5in]
        {\bf ABSTRACT}\\[.1in] \end{center} \begin{quotation}}  
\def\Title#1#2#3#4#5#6#7{\border\headpic
        {\hbox to\hsize{#7 \hfill UMDEPP #6}}\par
        \begin{center} \vglue .4in {\large\bf #1}\\[.4in]
        {#2}\\[.1in] {\it Department of Physics and Astronomy}\\
        {\it University of Maryland, College Park, MD 20742}\\[.1in]
        {#3}\\[.1in] {\it {#4}}\\ {\it {#5}}\\[.5in] {\bf ABSTRACT}\\[.1in]
        \end{center} \begin{quotation}}                 
\def\endtitle{\end{quotation}\newpage}                  

\begin{document}

\def\beq{\begin{equation}}
\def\eeq{\end{equation}}
\def\psib{\bar{\psi}}
\def\rh{\hat{{\bf r}}}
\begin{center}
\thispagestyle{empty}
{\bf \Large THE CONCEPT OF POTENTIAL IN QUANTUM FIELD THEORY}
\\
\baselineskip=12pt
\vspace{35pt}
J. Sucher

\vspace{24pt}

Center for Theoretical Physics and Department of Physics\\
University of Maryland, College Park, MD 20742\\
\vspace{40pt}

IN MEMORIAM

GERALD FEINBERG, 1933-1992
\end{center}
\vspace{40pt}

\begin{abstract}

	In quantum field theory the concept of a Lagrangian interaction density,
expressed in terms of fields, is primary.  Forces between two particles are
 regarded as arising primarily from the exchange of quanta of the bosonic
fields.  Thus,
in contrast to nonrelativistic quantum mechanics, the concept of a two-body
poten
tial is secondary.  Potentials are not given a priori but must be defined.
Part
of the purpose of this talk is to review and discuss the issues involved when
such
definitions are made.  In this context I describe a gauge-independent approach
to some aspects of the problem of determining the energy levels of bound
states,
developed in collaboration with the late G. Feinberg, and report some recent
results on the long-range potential associated with two-photon exchange between
charged particles.
\end{abstract}
\vfill

\baselineskip=24pt

* Invited talk, International Workshop on {\it Quantum Systems: New Trends and
 Methods }, Minsk, May, 1994.
\newpage
\addtocounter{page}{-1}
\section{INTRODUCTION}

The concept of potential is of course familiar to every physicist.  One first
encounters it in
 the context of nonrelativistic (n.r.) classical mechanics, after the
introduction of the
concepts of force and work.  For a point particle "1" moving in an external
conservative field,
 exerting a force which depends only on the position $\bf {r}_1$, there is
associated a function $V_1$
 of $\bf {r}_1$ whose negative gradient yields the force; $V_1$ is unique, up
to an additive constant.
  The associated classical Hamiltonian is then just $H(1) = K_1 + V_1$, where
 $K_1 = \bf {p}_1^2/2m_1$
 is the nonrelativistic kinetic energy.  Similarly, if the interaction between
particles
\lq \lq 1"
 and \lq \lq 2" is describable by a conservative force, dependent only on the
positions $\bf {r}_1$ and $\bf {r}_2$
,  the classical Hamiltonian is just $H(1,2) = K_1 + K_2 + V_{12}$, where the
potential
$V_{12} = V_{12}(\bf {r}_1,\bf {r}_2)$ is again uniquely determined by the
force, modulo a constant.
  The corresponding operator $H^{op}$ describing the dynamics of the system in
the context of
nonrelativistic quantum mechanics (NRQM) is given by making the replacements $
{\bf p}_i
\rightarrow  {\bf p}_i^{op} = -i\frac{\partial}{\partial {\bf r}_i}$ in
$H(1,2)$.  The best
known and physically most successful example of this procedure is familiar from
atomic physics,
 where the total potential operator $V^{op}$ is given by the sum of Coulomb
interaction potentials
$ U_C(i,j) = e_ie_j/4 \pi r_{ij}$, regarded as multiplicative operators in
coordinate space.

If the classical forces are velocity dependent, the situation is a bit more
complicated.  As a
 relevant example, consider $U_D$, the $v^2/c^2$ correction to the description
of the motion of
 two charged particles,
first obtained by Darwin from classical electrodynamics~\cite{r1},

	$$U_D = -\frac{1}{2c^2}({\bf {v}_1 \cdot \bf {v}_2}+
{\bf {v}_1 \cdot \bf {\hat{r}}\bf
{v}_2 \cdot \bf {\hat{r}}})U_C
 \, \, ~.\eqno(1.1)$$

On writing ${\bf v}_i = {\bf p}_i/m_i$ and then replacing $\bf {p}_i$ by ${\bf
 p}_i^{op}$ one obtains an operator $U_D^{op}$
which is however not unique because of the question of operator ordering.
Although, as will
be seen later, there is an ordering which is consistent with quantum
electrodynamics (QED),
the main point is that lack of uniqueness in potentials is the norm in
relativistic quantum
field theory (RQFT).

In RQFT the concept of a two-body potential (more precisely, of a potential
operator) is
 secondary.  What is primary is the concept of a Lagrangian interaction
density, expressed
 in terms of fields.  Potentials are not given {\it a priori} but must be
defined and their
use delineated.  The purpose of this talk is to review and discuss some of the
issues involved
when such definitions are made, to summarize some old results in this area, and
to report some
 ones, especially on the long-range potentials associated with the exchange of
photons between
charged particles, including spin-dependence.  By way of emphasizing the
importance of such an
 analysis, let us consider two examples, both of them involving spin-1/2
particles which in the
 absence of interaction are described by the free Dirac equation.

	i) The usual starting point for a relativistic theory of hydrogen or H-like
ions is
the external-field Dirac equation:

$$
h_{D;ext}(1)\Psi(1) \equiv [h_D^{op}(1)+V_{ext}(1)]\Psi(1) = E\Psi(1),
\hspace{1cm} [h_D^{op}(1)
 \equiv {\bf \alpha_1 \cdot p_1}+\beta_1 m_1]~\eqno(1.2)
$$
where $V_{ext}(1) = -Z\alpha/4\pi r_1$.  Now consider the equation
$$h(1,2)\Psi(1,2) \equiv [h_{D;ext}(1)+h_{D;ext}(2)+U_C]\Psi(1,2) = E\Psi(1,2)
\, \, ~,\eqno(1.3)
$$
with $U_C = k/r_{12}$ the $e^--e^-$ Coulomb interaction, long used as a
starting point for a
 relativistic theory of helium or He-like ions.  With regard to (1.3), we
have:\\
{\it Question 1}: Do you think that its use is (a) reasonable, (b)
unreasonable, or
that (c) the choice is not well posed.

ii) In the late 1920's G. Breit~\cite{r2}, using the then brand-new quantum
field theory (as
promulgated in a preprint of Heisenberg and Pauli!), and J. Gaunt,\cite{r3}
using analogy
 with classical electrodynamics, independently considered the question of the
leading
correction to the Coulomb interaction, in the context of a Dirac description of
electrons.
 However, they arrived at different results:
	$$U_B = -\frac{1}{2}({\bf \alpha_1 \cdot \alpha_2}+{\bf \alpha_1 \cdot
\hat{r}\alpha_2 \cdot
\hat{r}})U_C,  \hspace{2cm} U_G = -{\bf \alpha_1 \cdot \alpha_2}U_C
{}~.\eqno(1.4)                   $$
{\it Question 2}: Do you think that a) Breit was right, b) Gaunt was right, c)
both were right,
 or d) neither was right?  Explain your answer.

I will come back to these questions later.  The fact that they can be raised at
all
illustrates that in the context of RQFT the meaning of a two-body interaction
operator or
potential $V$ is not self-evident.  Indeed, the use of such potentials turns
out to be rather
 subtle, involving ambiguities and sometimes major traps \cite{r4}.
Historically,
effective potentials
have often \lq \lq emerged" in the context of level-shift calculations for a
specific physical system,
 initially in the context of ordinary time-independent perturbation theory,
later in the
 context of Tamm-Dancoff (TD) type of calculations, and still later from
four-dimensional
 Bethe-Salpeter (BS) type of equations; the latter associate a time $t_i$ with
each spatial
coordinate $\bf {r}_i$ and in practice it is necessary to carry out a messy and
approximate reduction
 to equal times \cite{r5}.  However, such potentials merit {\it a priori}
definitions and
careful delineation
 of their use.

One may ask why one should be interested in potentials at all in the context of
RQFT.
There are both practical and methodological reasons.  Practical, because
potentials have proved
 to be useful in studying a great variety of physical processes and systems at
low energies;
moreover, a huge amount of experience has been accumulated in dealing with the
solution of
 three-dimensional wave equations involving such potentials.  Methodological,
because in cases
 where the fundamental interaction is known, one ought to be able to explain
how the requisite
 potentials arise from the underlying theory in a clearcut, straightforward
way.

Suppose then that we wish to describe the interaction of
 particles in terms of potentials which can be used in 3-dimensional equations
and to define
 such potentials directly.  Obviously desirable criteria one might want to
impose on such
approach include:

  i) retaining the enormous simplification achieved by the use of Feynman
graphs and techniques
 in the computation of field-theoretic effects,

  ii) avoiding any {\it a priori} nonrelativistic approximations and, if
possible,

  iii) maintaining Lorentz and gauge invariance (GI) at any stage of
calculation.

Note that the BS equation involves a kernel $K$ which must be truncated in
practice; this
destroys GI in gauge theories such as QED.  Further, in bound-state problems
the use of Coulomb
 gauge is a practical necessity, which destroys manifest Lorentz invariance.
By way of
contrast, the approach I will sketch retains both Lorentz and gauge invariance
at any stage of
 approximation and, when restricted to the analysis of particle exchange, deals
only with
 on-shell Feynman amplitudes.  I believe it also has a higher ISQ (intellectual
satisfaction
quotient) than the traditional approaches, which to some extent have the
character of a black
box.  It was developed in collaboration with the late Gary Feinberg.

\section{S-MATRIX APPROACH TO EFFECTIVE POTENTIALS AND BOUND STATES}

Our approach has its genesis in work done long ago on the quantum theory of
long range forces
 (LRF).  Using the techniques of particle theory (Lorentz and gauge invariance,
analyticity and
 unitarity) we studied, in particular, the LRF arising from photon exchange
between two
composite neutral spinless systems \cite{r6} and later those between a charged
and a neutral
 \cite{r7} or another
 charged system \cite{r8}.
The basic idea is quite simple, a sort of a geometric mean between TD and BS
\cite{r8}.  Somewhat
 paradoxically, we first consider the scattering problem and the associated
two-body transition
 amplitude $T$.  We then ask to what extent T can be regarded as arising from
an effective
two-body potential, to be used in a Schr\"{o}dinger type of equation.  To be
more explicit, we
define an interaction operator $V$, acting directly in configuration space, as
a Fourier
transform of an on-shell amplitude, obtained from gauge-invariant subsets of
Feynman diagrams,
 modified by appropriate subtractions to avoid double counting; $V$ is
constrained by the
requirement that when used in a specified type of relativistic Schr\"{o}dinger
equation it
 reproduces $T_{c.m.}$, the value of $T$ in the center-of-momentum system
(c.m.s.).
For spin-0 particles this equation is taken to have the natural form
$$
h\phi = W\phi, \hspace{3cm} h = h_o + V ~\eqno(2.1{\rm a})$$
with $h_o$ defined by
	$$h_o = E_A^{op} + E_B^{op} , \hspace{2cm} [E_i^{op} \equiv \sqrt{ m_i^2+{\bf
p}_{op}^2},\,
{\bf p}_{op} = -i\partial / \partial {\bf r}]  \, \, ~.\eqno(2.1{\rm b})$$
The associated potential-theory transition amplitude $T_{pot}$ is given by
$$
	T_{pot} = \langle {\bf p}'| V + V(W-h_0-V+i\epsilon)^{-1}V|{\bf p}\rangle \,
\, ~.\eqno(2.2)$$
The field-theory transition amplitude $T$ is given in the c.m.s. by
$$
	T_{c.m.} = M(s,t)/4 E_A E_B ~\eqno(2.3{\rm a})$$
where $M(s,t)$ denotes the invariant Feynman amplitude and $s$ and $t$ are the
invariant
 squares of energy and momentum transfer, respectively:
$$
s \equiv (p_A+p_B)^2, \hspace{3cm} t \equiv (p_A-p_A')^2 \, \, ~.\eqno(2.3{\rm
b})$$
The potential $V$, which in general will be nonlocal and/or depend
parametrically on $s$, is
then required to generate $T_{c.m.}$ from (2.2),
$$
	T_{pot} = T_{c.m.} \, ~,\eqno(2.4)$$
a condition which is to be satisfied order-by-order in perturbation theory.  To
apply this to
 bound states, we look for normalizable solutions of (2.1).  The associated
eigenvalues will
correspond to poles of $M$ at values of $s$ below the threshold $ s_0 =
(m_A+m_B)^2$ and so can be
 interpreted as the masses of bound states.

\section{USE OF ANALYTICITY AND UNITARITY}

In the computation of potentials from scattering amplitudes it is both
convenient and
 physically appealing to utilize the fact that such amplitudes are analytic
functions of the
 variables on which they depend.  In particular, the  contribution $M_S(s,t)$
to $M$ from a
set $S$ of Feynman diagrams is, for fixed $s$, usually found to be an analytic
function of $t$,
 now
regarded as a complex variable, with singularities only on the real t-axis.  If
$M_S$
vanishes as $t \rightarrow \infty$, one can use Cauchy's theorem to write
$M_S(s,t)$ in the
form
$$
	M_S(s,t) = \pi^{-1} \int dt' \rho_S(s,t')/(t'-t) \, \, ,\eqno(3.1{\rm a})$$
where $\rho_S(s,t)$, the so-called spectral function, is proportional to the
discontinuity of
 $M_S(s,t)$ across the real t-axis:
	$$\rho_S(s,t) = (2i)^{-1}[M_S(s,t+i0)-M_S(s,t-i0)] ~.\eqno(3.1{\rm b})$$
A practical advantage of this relationship is that the spectral function is
often relatively
easy to calculate and/or expressible in terms the amplitudes associated with
other physical
processes, by use of the ideas of unitarity or generalized unitarity \cite{r9}.
  A
conceptual advantage
 is that ambiguities in the potential $V_S$ corresponding to $M_S$ may be
limited by a
sensible requirement: It should be defined in such a way that is reproduces
$M_S(s,t)$ not only in
the physical region of the scattering, i.e. for
$$
     s \ge s_0 \equiv (m_A+m_B)^2 , \hspace{3cm} -4p^2 \le t \le 0 \, \,
,\eqno(3.2)$$
with $p$ the magnitude of the 3-momentum of either $A$ or $B$ in the c.m.s.,
but also for $t$ outside
this region, where $M_S(s,t)$ is uniquely determined by analytic continu-ation.
 After all, the
analyticity properties of $M_S(s,t)$ arise from the deepest properties of RQFT,
namely locality,
 and reproducing the properties of field theory is the leitmotif of our
approach.  For spin-0
particles it is useful to eliminate the kinematic energy factors in (2.3) by
defining a
modified potential $U$ via
$$
	V = y_{op}Uy_{op}\, ,\hspace{3cm}  y_{op} \equiv
\sqrt{m_Am_B/E_A^{op}E_B^{op}} ~\eqno(3.3)$$
and to require that $U$ be local, depending only parametrically on $s$ or
equivalently on $p^2$.
We then find, by inversion of the Fourier transform and use of the spectral
representation
(3.1), that $U_S = U_S(r;p^2)$ may be expressed directly in terms of $\rho_S$:
$$
U_S(r;p^2) = (16\pi^2 m_Am_B)^{-1} \int dt \rho_S(s,t) \exp(-\sqrt{t}r)
{}~.\eqno(3.4)$$

In general, there is a nearest right-hand branch point at a value $t_0 \ge 0$,
equal to the
 minimum mass of the particle systems being exchanged by A and B in the graphs
included in the
 set $S$, and a nearest left-hand branch point at a value $\bar{t}_0 < 0$; the
function $\rho_S$
 vanishes in the interval $(\bar{t}_0,t_0)$.  It can be shown that the
contribution to $U_S$
 from the region $t \le \bar{t}_0$ always gives rise to a short-range
potential, i.e. one
which vanishes exponentially as $r$ becomes large \cite{r8}.  However, if $t =
0$, as
 is always the case when only zero-mass quanta such as photons or neutrinos are
exchanged, the
 integral from $0$ to $\infty$ yields a long-range (LR) potential $U_S^{LR}$,
i.e. one which falls
off as an inverse power of $r$ for large $r$:
$$
	U_S^{LR}(r;p^2) =(16\pi^2 m_Am_B)^{-1} \int_{0}^{\infty} dt \rho_S(s,t)
\exp(-\sqrt{t}r) .\eqno(3.5)$$
Thus the relation (3.4) is especially convenient for the analysis of a
long-range
potential (LRP), associated with the exchange of zero-mass quanta.  For it is
clear from the
Laplace transform character of (3.5) that to determine the asymptotic form of
$U_S^{LR}$ at large $r$
 it suffices to know the behavior of $\rho_S$ near $t = 0$.

For the physical examples which have been studied for the $t_0 = 0$ case, the
spectral function
 $\rho$ can be represented in the neighborhood of $t = 0$ by a Laurent
expansion in $\sqrt{t}$,
 with a simple pole. Dropping the subscript $S$, we have \cite{r8}
$$
	\rho(s,t) = a_2(s)t^{-1/2} + a_3(s) + a_4(s)t^{1/2} + \ldots .\eqno(3.6)$$
Substitution into (3.5) then yields an expansion for ULR in inverse powers of
$r$, with
coefficients which depend on $s$, or equivalently on the c.m. momentum $p$:
$$
	U^{LR}(r;p2) = c_2(p^2)r^{-2} + c_3(p^2)r^{-3} + c_4(p^2)r^{-4} +
\ldots,\eqno(3.7{\rm a})$$
where
$$c_n(p^2) = [(n-2)!/8\pi^2 m_A m_B]a_n(s) \, \, .\eqno(3.7{\rm b})$$
With this as background, we are ready to turn to some specific applications.

\section{POTENTIALS FROM ONE- AND TWO-QUANTUM EXCHANGE}

	Let us consider some examples of forces arising from the exchange of one or
two quanta
between spin-0 particles A and B, first for massive quanta and then for
massless quanta.

{\bf A. Exchange of massive spin-0 quanta}

As perhaps the simplest example of the techniques sketched above, consider a
theory of two
complex scalar fields $\phi_A$ and $\phi_B$, both interacting with a neutral
scalar field $\phi$
 of mass $\mu$,
 with an interaction Lagrangian  ${\cal L}_I = -G_A \phi_A^{*}\phi_A \phi +
(A\rightarrow B)$.
 The lowest order Feynman amplitude is
then given by $M^{(2)} = G_AG_B(t-\mu^2)^{-1}$; there is now a pole instead of
a branchpoint,
 but the
formalism still applies.  The (formal) discontinuity of $M^{(2)}$ is $-2\pi i
G_AG_B  \delta (t-
\mu^2)$ and the
 spectral function is $-\pi G_AG_B \delta (t-\mu^2)$.  Thus, with $g_i \equiv
G_i/2m_i$ the
 associated dimensionless
 coupling constants, (3.5) yields
$$
	U^{(2)} = U_Y \equiv -(g_Ag_B/4\pi r)exp(-\mu  r) \, \, .\eqno(4.1)$$
This is of course just the potential first obtained by Yukawa, for infinitely
massive nucleons.
The corresponding $V^{(2)} = y_{op}U_Yy_{op}$ takes into account recoil
corrections to the
static potential
 $U_Y$, to all orders in $v/c$.  Note that $U^{(2)}$ is
independent of $s$ only because $M^{(2)}$ is.  In higher orders this feature
disappears.
	A more complicated example in the same theory is provided by consideration of
the fourth-order potential arising from the exchange of two quanta.  There are
now two Feynman
 diagrams, a two-rung ladder or ``box" diagram with amplitude $M_a^{(4)}$ and a
``crossed box" diagram
with amplitude $M_b^{(4)}$, whose sum may be written in the form
$$
M^{(4)} =i\frac{G_A G_B}{2}\int d^4 k d^4 k'  \delta
(Q-k-k')\frac{(D_A^{-1}+D_A'^{-1})(D_B^{-1}+D_B'^{-1})}{
(k^2-\mu^2+i\epsilon)(k'^2-\mu^2+i\epsilon)} \eqno(4.2)$$
with $ D_A = (p_A-k)^2-m_A^2+i\epsilon$, $D_B = (p_B+k)^2-m_B^2+i\epsilon$ and
$D_A'$, $D_B'$
 obtained by letting $k  \rightarrow k'$ in $D_A$, and
 $D_B$.  $M^{(4)}$ has a branchpoint at $t_0 = 4\mu^2$ and the discontinuity
across the cut
extending from $t_0$
to $+\infty$ is obtained by using the generalized unitarity theorem of Cutkosky
\cite{r9}, which
requires
replacing each meson propagator by its discontinuity:
$$
	(k^2-\mu^2+i\epsilon)^{-1}-(k^2-\mu^2-i\epsilon)^{-1} \rightarrow -2\pi i
\delta(k^2-
\mu^2)   \eqno(4.3)$$
and including a factor $\theta(k^0)$ to insure that the mesons have positive
energy.

	As mentioned before, to avoid double counting one must subtract from
$M^{(4)}$ the iteration
 amplitude $M_I^{(4)}$, obtained from $V^{(2)}$ in second-order perturbation
theory, before
computing the
 potential.  This amplitude is also analytic in $t$ with a cut at $t_0$ and its
discontinuity
 may be
 calculated directly.  One obtains in this way a ``net" fourth-order spectral
function:
$\rho_{net}^{(4)}
 = \rho^{(4)}-\rho_I^{(4)}$, which now depends not only on $t$ but also on
$p^2$.  For $p^2 = 0$,
one finds, using
 (3.4), that \cite{r10}
$$
	U_{net}^{(4)} = (m/\mu)^2(g_Ag_B/4\pi r)^2 \exp(-2\mu r)[1-\eta(\mu r)^{-1/2}
+ O(r^{-1})] \eqno(4.4)$$
where $\eta = 4m/\mu \sqrt{\pi}$.  For $ \mu /M \approx 1/7$, the pion-nucleon
mass ratio, the
factor $\eta$ is quite large,
 about 15. Thus in this case the $r^{-5/2}$ term, coming from $M^{(4)}$,
dominates the
${\bf r}^{-2}$ term, coming
from $M_I$, for distances of the order of 10 fermi or less.

{\bf B. Exchange of photons}

	Now let us consider the LRF arising from exchange of photons between spinless
systems $A$
 and $B$.  If at least one of $A$ and $B$ is neutral, the one-photon exchange
potential $V_{1\gamma}$
 is short-range.  However, the two-photon exchange potential $V_{2\gamma}$ can
be long-range.
 Further, in
studying the large-$r$ behavior of $V_{1\gamma}$ we need not worry about the
effects of iteration of
$ V_{1\gamma}$,
 since this can only contribute to short-range effects.  So this case is in
some ways actually
simpler than that of two charged particles and I consider it first.
	In the study of $V_{2\gamma}$ a key role is played by the amplitude for photon
scattering by
either particle.   As a consequence of Lorentz invariance, for a spinless
particle this
amplitude has the generic form
$$
	M(\sigma,t) = M^{\mu \nu}(p',k';p,k)\epsilon_{\mu}\epsilon_{\nu}'^{*} \,
.\eqno(4.5)$$
Using gauge invariance one can show that on the photon mass-shell the tensor
amplitude
$M^{\mu \nu}$ may
 be written as
$$
\label{e-46}
	M^{\mu \nu}(p',k';p,k) = F_E(\sigma,t)T_E^{\mu \nu} + F_M(\sigma,t)T_M^{\mu
\nu} \, ,\eqno(4.6{\rm a})$$
where the invariant amplitudes $F_E$ and $F_M$ are functions only of $\sigma$
and $t$, the
 invariant squared energy and squared momentum transfer,
$$\sigma = (p+k)^2, \hspace{1cm} t = (p-p')^2,\hspace{1cm}
\hspace{1cm} \bar{\sigma} = (p-k')^2 = -\sigma-t+2m^2 \eqno(4.6{\rm b})$$
The quantity $\bar{\sigma}$, the cross-momentum transfer, is defined for use
below. The notation corresponds
to a special choice of gauge-invariant tensors $T_i^{\mu \nu}$ which may be
regarded as
\lq \lq electric " and  \lq \lq magnetic", but which I need not reproduce here.
Their main feature is that {\it if the particle is neutral}, the accompanying
coefficients may be
 shown to have the property
$$
	F_i(m^2,0) = 4\pi \alpha_i \, ,\eqno(4.7)$$
where $\alpha_E$ and $\alpha_M$ denote the static electric and magnetic
polarizabilities of the
 particle.
This explains the nomenclature.  Moreover, the $F_i$ admit spectral
representations of the form
$$
	F_i(\sigma,t) = \pi^{-1} \int d\sigma' \rho_i(\sigma',t)
[(\sigma'-\sigma)^{-1}+ (\sigma'-\bar{\sigma})^{-1}] \, \, .\eqno(4.8)$$
where $\bar{\sigma}$ is defined by (4.6b).  With $\omega \equiv
(\sigma-m^2)/2m$, one finds from (4.7) and (4.8) the sum rule
$$
	2\pi^2 \alpha_i = \int_{0}^{\infty} d \omega \rho_i(\sigma,0)/\omega \,
.\eqno(4.9)$$
In terms of such tensors, the amplitude for two-photon exchange is given by
$$
	M_{2\gamma}^{(4)} = (i/2)\int \int d^4 kd^4 k' \delta (Q-k-k')\frac{
M_A(p_A',k';p_A,-k):M_B
(p_B',-k';p_B,k)}{(k^2-\mu^2+i\epsilon)(k'^2-\mu^2+i\epsilon)} \,
.\eqno(4.10)$$
Here $M_A^{\mu \nu}(p_A',k';p_A,-k)$ is the tensor amplitude for the emission
of two virtual
 photons by $A$
and $M_B^{\mu \nu}(p_B',-k';p_B,k)$ that for absorption of two virtual photons
by $B$, both of
 them off-shell
 extensions of the on-shell tensors; the colon in (4.10) denotes a summation
over tensor
indices.

{\bf 1. Both A and B neutral}

	When one takes the t-discontinuity of $M_{2\gamma}$ by using (4.3), the
photons go on the mass
 shell and, with both $A$ and $B$ neutral, one can use the form (4.6) for both
$M_A^{\mu \nu}$
 and $M_B^{\mu \nu}$ in
(4.10) to carry out the indicated contraction.  This already shows that
$U_{2\gamma}$ is a quadratic
functional of the invariants $F_i^A$ and $F_i^B$.  To obtain the asymptotic
form of $U_{2\gamma}$
 one needs
only the value of the spectral function $\rho_{2\gamma}$ near $t = 0$ and this
can be expressed in
 terms of
 integrals involving the spectral functions $\rho_i^A(\sigma,0)$ and
$\rho_i^B(\sigma,0)$.
 On use of the sum rule
(4.7) one finds that for large $r$ and low energies $V_{2\gamma}$ falls of as
$r^{-7}$, with a
 coefficient
which is a quadratic function of the static polarizabilities \cite{r6},
$$
	V_{2\gamma} \approx -D/{\bf r}^7, \hspace{1cm} D \equiv (23/4\pi)(\alpha_E^A
\alpha_E^B+\alpha_M^A
\alpha_M^B)-(7/4\pi)(\alpha_E^A \alpha_M^B+\alpha_M^A \alpha_E^B) \,
.\eqno(4.11)$$
The purely electric terms were first obtained by Casimir and Polder \cite{r11}.

In the case of two atoms, (4.11) is a good approximation only for separations
which
are large compared to the maximum wavelengths for dipole emission, of
order $\alpha^{-1}a$
 with $a$ the
 Bohr radius.  The potential between two atoms at smaller distances, but still
large compared
to a, can also be studied by these methods \cite{r6}  The form of the integrals
involved suggested that
 for $r$ large compared to $a$, a very good approximation to $V_{2\gamma}$
would be given by an arctangent
 function \cite{r12},
$$
	V_{2\gamma} \approx -(C/r^6)(2/\pi)\arctan(d/r),\eqno(4.12)$$
where $d = (23/8C)\alpha_A \alpha_B$ and $C$ is the so-called London constant.
Comparison
with available
numerical calculations shows that (4.12) interpolates with better than
two percent accuracy
 between the
 Wang-London \cite{r13} potential $V = -C/r^6$, which neglects retardation, and
the
 asymptotic formula
 (4.11) with magnetic effects neglected.

{ \bf 2. A neutral, B charged}

The same techniques can be applied to the case of a neutral composite $A$ and a
charged
particle $B$, since $V_{1\gamma}$ is still short-range \cite{r7}.  However, the
invariant
amplitudes $F_i^B$ then
 contain pole terms and the identification (4.7) fails.  For a an elementary
$B$, with charge
$e_B$,
 these pole terms can be calculated explicitly by using scalar QED; they
correspond to
 contributions to the spectral function $\rho_B$ which are proportional to
$\delta(\sigma - m_B^2)$
.  The result is
$$
	V_{2\gamma}(r) =
-(e_B^2/4\pi)[(1/2)\alpha_Er^{-4}-(11/4\pi)\alpha_Er^{-4}(\lambda_B/
 r)-(5/4\pi)\alpha_M r^{-4}(\lambda_B/r)+\ldots]\eqno(4.13)$$
where $\lambda_B = m_B^{-1}$, the $\alpha$'s refer to the polarizabilities of
$A$, and the
 dots denote terms which
 fall off as $1/r^7$ or faster.  The $11/4\pi$  term was first found by J.
Bernabeu and
 R. Tarrach \cite{r14},
 using the present methods, and by E. Kelsey and L. Spruch \cite{r15}, using
hybrid QED.  The latter
 authors also suggested that its presence could be tested by study of the fine
structure of
 Rydberg states of helium.  Measurements of these were carried out by S.
Lundeen and coworkers
 over a period of years, with $n = 10$ for the outer electron \cite{r16}.  It
turns out that the
 asymptotic formula (4.13) is not accurate enough at $n = 10$, but a general
theory based on the
 present method can be worked out which gives the potential at any separation
large compared
 to the Bohr radius  \cite{r7} \cite{r17} \cite{r16}.  The whole subject is
discussed at length in Ref. 16, which
 includes articles by Spruch, G.W. F. Drake, R. Drachman, and by Feinberg and
me.

\section{  LRF BETWEEN CHARGED PARTICLES: BEYOND THE COULOMB POTENTIAL}

	The extension of these methods to the case of two charged particles runs at
once into
 a serious difficulty: Some of the integrals associated with two-photon
exchange are infrared
IR) divergent.  Thus it appears at first sight that a two-photon exchange
potential does not
 exist!  Some reflection leads to the realization that these IR divergences are
the counterpart
 in quantum field theory of the well-known fact that in NRQM the Coulomb
interaction cannot be
 treated in perturbation theory; the second Born approximation diverges, not
just for zero
 momentum transfer but for any value of $|t|$.  The cure for this problem turns
out to be
 precisely the subtractions which are necessary to avoid double counting
\cite{r8}.

	For concreteness, let us study the case of two point-like spin-0 particles,
with
 charges $e_A$ and $e_B$, and confine our attention to the so-called
generalized ladder approximation
 to $M(s,t)$, i.e. to graphs which only involve photon exchange {\it between}
the particles.

{\bf A. One-photon exchange potential}

	Before considering two-photon exchange we must define a one-photon exchange
potential
 $V_{1\gamma}$.  Note that however $V_{1\gamma}$ is defined, it must reduce to
the Coulomb
potential in
 the static
limit.  Since this is long-range, the associated iteration amplitude $M_I$ is
likely to be
 equivalent to a long-range potential.  Thus, even if one were unaware of the
IR divergence
problem one would have to compute MI to find  just the long-range part of
$V_{2\gamma}$.

	If one uses Feynman gauge in writing down the (gauge invariant) one-photon
exchange
amplitude $M_{1\gamma}$, one gets, on use of the relation $(p_A+p_A')\cdot
(p_B+p_B') = s-u$,
with \linebreak
$ u \equiv (p_A-p_B')^2 $
the cross momentum transfer, $M_{1\gamma} = e_Ae_B(s-u)/t$.  Since \linebreak
$u = 2m_A^2+2m_B^2-s-t$ we have
$$
	M_{1\gamma} = e_Ae_B(2a+t)/t   \hspace{2cm} (a \equiv s-m_A^2-m_B^2) \, \, .
\eqno(5.1)$$
Simple Fourier transformation of (5.1) yields a term proportional to $U_C$,
with an
 energy-dependent coefficient, plus a contact term proportional to $\delta({\bf
r})$.  Such a
potential is
 not suitable for use in a Schr\"{o}dinger type of equation.  In second-order
perturbation theory it
would lead to an ultraviolet ($UV$) divergence.  A potential which is iterable
can be obtained by
 first writing $M_{1\gamma}$ in a different form (which does not change its
value on the mass shell) and
 then finding an equivalent operator in r-space which involves derivative
operators \cite{r8}.  One is
thereby led to what can be termed a Feynman-gauge-inspired (FGI) potential
$V_{1\gamma}^{FGI}$,
$$
	V_{1\gamma}^{FGI} =  z'_{op}U_C z'_{op} + y_{op}({\bf p}_{op} \cdot U_C {\bf
p}_{op}/2m_Am_B)
y_{op},\eqno(5.2)$$
where $z'_{op} \equiv \sqrt{1+p_{op}^2/2E_A^{op}E_B^{op}}$ and $y_{op} = \sqrt{
m_Am_B/E_A^{op}
E_B^{op}}$.  The corresponding
Coulomb-gauge inspired (CGI) one-photon exchange potential $V_{1\gamma}^{CGI}$
is given by
\cite{r18}
$$
	V_{1\gamma}^{CGI} \equiv y_{op}[\{ E_A^{op},\{ E_B^{op},U_C\}  \}+(1/2)
\{{\bf p}_i^{op},\{ {\bf p}_j^{op},(\delta_{ij}+\hat{{\bf r}}_i\, \hat{{\bf
r}}_j)U_C\} \}]
y_{op}/4m_Am_B \, .\eqno(5.3)$$
The curly brackets denote anticommutators.  The $t$ discontinuity of the
iteration amplitude
$M_I$
obtained from either choice is IR finite but behaves as $1/t$ for small $t$;
this behavior leads
to a logarithmic divergence in the spectral integral, consistent with the
nature of the IR
divergence of $M_I$ itself.\\
{\bf B. Two-photon exchange potential}

To compute the field theory amplitude $M^{(4)}$ in scalar QED one must study
the integrals
associated with the five fourth-order Feynman diagrams which enter the game:
(a) the two-rung
ladder graph, (b) the two-rung crossed ladder graph, (c) the two single-seagull
graphs and (e):
 the double-seagull graph.  Both (a) and (b) are UV convergent but IR
divergent, whereas (c)
 and (d) are UV divergent but IR convergent.  But the $t$ discontinuity of each
of these is
divergence free.  The net spectral function behaves again as $1/t$ for small
$t$, corresponding
to
the IR divergence but the coefficient of $t^{-1}$ is equal and opposite to that
appearing in
$M_I$.
The difference spectral function $\rho_{diff}$ then goes like $t^{-1/2}$, which
is integrable at
$t = 0$ and
 yields a cutoff-independent $V_{2\gamma}$.  On using $V_{1\gamma}^{FGI}$ to
compute $M_I$ one finds
that\cite{r8}
$$
 	V_{2\gamma}^{FGI} = c_2^{FGI}r^{-2} + c_3^{FGI}r^{-3} + \ldots\eqno(5.4{\rm
a})$$
where, with $k \equiv e_Ae_B/4\pi$,
$$c_2^{FGI} = k^2/2(m_A+m_B), \hspace{2cm} c_3^{FGI} = - 7k^2/6\pi m_Am_B \, \,
.\eqno(5.4{\rm b}$$
In contrast, use of $V_{1\gamma}^{CGI}$ yields \cite{r18}
$$
  	V_{2\gamma}^{CGI} = c_2^{CGI}r^{-2} + c_3^{CGI}r^{-3} + \ldots \eqno(5.5{\rm
a})$$
where
$$c_2^{CGI} = 0, \hspace{2cm} c_3^{CGI} = - 7k^2/6\pi m_Am_B \, \,
.\eqno(5.5{\rm b})$$
Thus we see that in the case of two charged particles the leading asymptotic
behavior of
$V_{2\gamma}$
depends on the precise definition of $V_{1\gamma}$.

	This observation resolves a long-standing puzzle in the literature and shows
that in
the case of two charged particles not just the concept of potential but even
that of its
asymptotic form is not without ambiguity.  Further, as was noted some time ago
by L. Spruch,
$c_2^{FGI}$ is classical in character, i.e. if $h$ and $c$ are restored, $c_2$
turns out to be
 independent
of $h$.  One should therefore try to understand the source of this term from
classical
electrodynamics.  It turns out that this is indeed possible, on the basis of a
reexamination
 of the pioneering work of Darwin on the effective Lagrangian describing the
interaction of
two charged particles to order $1/c^2$, but I will not enter into the details
here \cite{r19}.

{\bf C. Orbit-orbit potential}

	It is of some interest to study the difference between the two choices of
$V_{1\gamma}$.
It turns our that this is connected with the form of the so-called orbit-orbit
interaction.
To see this, note that in the n.r. limit (5.2) yields as the leading correction
to $U_C$ an
orbit-orbit interaction $U_{o-o}$ of the form
$$
U_{o-o}^{FGI} = \{ {\bf p}_i^{op},\{ {\bf p}_j^{op},\delta_{ij}U_C\} \}/4m_Am_B
\, ,\eqno(5.6)$$
whereas (5.3) yields
$$
	U_{o-o}^{CGI} = (1/2)\{ {\bf p}_i^{op}, \{ {\bf p}_j^{op},(\delta_{ij}+
\hat{{\bf r}}_i \, \hat{{\bf r}}_j)U_C \} \} /4m_Am_B \, \, .  \eqno(5.7)$$
The latter is a manifestly hermitian form of the orbit-orbit interaction
$U_{o-o}$ familiar from
atomic physics. It is usually obtained by reduction of the Breit operator (1.4)
to n.r. form;
this is unfortunate from a pedagogical point of view since, as one expects on
classical
grounds and as the calculation confirms, spin has nothing to do with it.
	The difference between these two forms of $U_{o-o}$ is precisely accounted for
by the
 $1/r^2$
term.  In the context of two-body Coulomb bound states, one can show that if
one starts with
the n.r. Schr\"{o}dinger equation
$$
	h_{nr}\phi = ({\bf p}_{op}^2/2m_{red})\phi + U_C\phi = W\phi \, ,\eqno(5.8)$$
the leading term $\delta W^{(4)}$ in the level shift $\delta W$ associated with
electromagnetic effects, of order
 $\alpha^4m_{red}$, is given by
$$
	\delta W^{(4)} = \langle \phi |U_{o-o}^{CGI} |\phi \rangle = \langle \phi
|U_{o-o}^{FGI}
 + c_2^{FGI}/r^2 |\phi \rangle\eqno(5.9)$$
The equivalence of these two expressions can be checked most easily by showing
that the difference betweeen the operators in question has the form of a
commutator $[h_{nr},X]$.
	This is an example of the fact that a major source of non-uniqueness of
two-body
potentials4 is the possibility of adding a term of the form $[h_0,X]$ to the
lowest-order
potential without changing the lowest-order scattering amplitude.

\section{INCLUSION OF SPIN-1/2}
{\bf A. Continuum dissolution}

The inclusion of spin-1/2 particles is straightforward, once one recognizes the
 main pitfall encountered when dealing with relativistic Dirac-like equations.
Consider, for
example, the helium atom or a He-like ion.  A natural starting point for a
potential theory
description of this system, which avoids making nonrelativistic approximations,
would appear to
be Eq.(1.2):
$$
	h(1,2)\Psi(1,2) \equiv [h_0(1,2)+U_C]\Psi(1,2) = E\Psi(1,2) \, \,
.\eqno(6.1{\rm a})$$
with
$$h_0(1,2) = h_{D;ext}(1)+h_{D;ext}(2) \, .\eqno(6.1{\rm b}))$$
This equation, and its obvious generalization to more than two electrons, has
appeared in the
literature for over fifty years, since first used by Breit around 1930.
However, it has no
normalizable solutions which correspond to the discrete spectrum of the atom or
ion, as first
pointed out by Brown and Ravenhall  \cite{r20}.  To see this, note that if
$U_C$ is dropped, the spectrum
of the residual Hamiltonian is the whole real line, with every point
nondenumerably degenerate;
 the degeneracy arises from the fact that if we consider a product
eigenfunction of $h_0(1,2)$
with one electron in a positive-energy continuum state and the other electron
in a
negative-energy state, we can increase the positive energy and decrease the
negative energy
continuously without changing their sum.  Thus, any product bound state of
$h_0(1,2)$ is
 embedded
in a sea of non-normalizable states. The turning on of any local interaction
will then lead to
a dissolving of this state into the continuum.  The correct answer to {\it
Question 1} is
therefore
 (b).  The cure for this disease, which I like to call ``continuum
dissolution", is to go back
to first principles.  Using field theory, one finds that putative interaction
potentials such
 as $U_C$ always come accompanied by positive-energy projection operators which
keep the
product bound states from mixing with the states responsible for the continuous
degeneracy.
Thus, in QED one finds that a suitable starting point for a relativistic theory
of He-like ions
 is provided by the "no-pair" Hamiltonian $H_{++}$ defined by
$$
	H_{++} = h_0(1,2)+L_{++}U_CL_{++}\eqno(6.2)$$
where $L_{++}$ is a product of external-field positive-energy projection
operators \cite{r21}.
	In the absence of an external field, i.e. in the pure two-body case, the
problem is
ameliorated because of momentum conservation.  However, starting from field
theory one again
finds that the effective interaction operators come equipped with projection
operators for the
 Dirac particles.  In particular, for two spin-1/2 particles the counterpart of
Eq.(6.2)
 is
$$
	[h_A^{op}+h_B^{op}+\Lambda_{++}^{op}U\Lambda_{++}^{op}]\Psi = E\Psi \, \, ,
\eqno(6.3)$$
where $h_A^{op} = {\bf \alpha}_A \cdot {\bf p}^{op}+\beta_Am_A$,  $h_B^{op} =
 -{\bf \alpha}_B \cdot {\bf p}^{op}+\beta_Bm_B$, and $\Lambda_{++}^{op}$ is the
product of free
 (Casimir-type)
 positive-energy projection operators:
$$
	\Lambda_{++}^{op} = \Lambda_{+;A}^{op}\Lambda_{+;B}^{op}, \hspace{2cm}
\Lambda_{+;i}^{op} = (E_i^{op}+h_i^{op})/2E_i^{op} \, \, .\eqno(6.4)$$
{\bf B. One- and two-photon exchange potentials}

	With this understanding, the one-photon exchange potential takes the form
$$
	V_{1\gamma} = \Lambda_{++}^{op}U_{1\gamma}\Lambda_{++}^{op} \, ,\eqno(6.5)$$
where $U_{1\gamma}$ is required to reproduce $M_{1\gamma}$ when sandwiched
between on-shell
Dirac spinor
 plane
 waves.  This leads to two natural choices for $U_{1\gamma}$ which are local in
Dirac-spinor
 space,
$$
U_{1\gamma}^{FGI} = U_C+U_G, \hspace{2cm}  U_{1\gamma}^{CGI}= U_C+U_B \,
,\eqno(6.6)$$
where $U_G$ and $U_B$ are the Gaunt and Breit potentials defined by Eq.(1.4).
It follows
 that,
strictly speaking, the answer to {\it Question 2} in Sec.I is (d).  The
projection operators are
 vital if the two-body equation is to have a natural origin in field theory and
if the equation
 is to lend itself to obvious generalization to more than
two particles.  They arise from the fact that, in contrast to Dirac's
one-electron theory, in
hole theory transitions to the filled negative-energy sea are forbidden by the
exclusion
 principle, a feature which is incorporated by QED.

	The computation of $V_{2\gamma}$ for the case when both particles have
spin-1/2 is a
 major
undertaking.  If one uses $V_{1\gamma}^{FGI}$ to compute $M_I$, the
spin-independent part (more
 precisely, the
 Dirac-matrix independent part) of $U_{2\gamma}$ is the same as that found for
two spin-0
particles.
 For the mixed case of, say, $A$ with spin-0 and $B$ with spin-$1/2$, the
computation of the
spin-dependent part has already been carried out \cite{r22}.  One finds a
correction
$V_{2\gamma}^{s-o}$
 to the
spin-orbit potential $V_{1\gamma}^{s-o}$ coming from $V_{1\gamma}$, which is
proportional to
${\bf \sigma \cdot l} /r^4$ for large $r$.
If $A$ has structure, $V_{2\gamma}^{s-o}$ also contains a spin-orbit
polarizability potential falling of as
${\bf \sigma \cdot l}/r^6$ for large $r$.  These terms will also be present in
the case of two
 spin-1/2 particles
 but in addition there will be a correction $V_{2\gamma}^{s-s}$ to the
spin-spin potential
$V_{1\gamma}^{s-s}$ coming
from $V_{1\gamma}$.  The calculation of this is underway in collaboration with
G.
Gilbert~\cite{r23}.

	There are a number of physical situations in which it may be possible to
detect the
effects of $V_{2\gamma}^{s-o}$.  Typically these involve measurements of bound
state energies in
exotic
 atoms, where one particle has spin 1/2 and another has spin 0.  Examples
include anti-protonic
atoms with a spin-0 nucleus, such as p-$\mbox{He}^4$, pionic atoms with a
spin-1/2 nucleus,
such as
 pionic hydrogen, and the pi-muon bound state known as pi-muonium.  Certain
aspects of
$V_{2\gamma}^{s-o}$
may be observable in Rydberg states of helium-like ions whose nuclei have spin
1/2.  For
 details see Ref. 22.\\
{\bf C. The two-neutrino exchange force.}

	As a final example, let us study the LRP arising from the exchange of
0-massless
spin-1/2 quanta.  A physical case is provided by the exchange of neutrinos
between two
 spin-1/2 particles \cite{r24}.  Although one-quantum exchange is forbidden by
angular momentum
conservation, the exchange of a neutrino-antineutrino pair is not. Consider a
current-current
 interaction of Dirac fields $\psi_A$ and $\psi_B$ with a massless neutrino
field $\psi_{\nu}$
 of the form
$$
	{\cal L}_{eff} = - G_A(\psib_A \Gamma_A^{\sigma}
\psi_A)(\psib_{\nu}\Gamma_{\sigma}
\psi_{\nu}) + (A \rightarrow B)\, ,\eqno(6.7)$$
where the $\Gamma$'s have the generic form
$$
	\Gamma_A^{\sigma} = \gamma_A^{\sigma}(1+\zeta_A \gamma_A^5), \hspace{1cm}
\Gamma_B^{\sigma} =\gamma_b^{\sigma}(1+\zeta_B \gamma_B^5),
\hspace{1cm}\Gamma_{\sigma}
 = \gamma_{\sigma}(1+c\gamma^5) \eqno(6.8)$$
and the $G$'s denote effective Fermi coupling constants.  One then finds that
the long-range
 part of the potential arising from exchange of such a pair is given by
\cite{r25}
$$
	V_{2\nu} = (G^2/4\pi^3)(\gamma_A^0 \gamma_B^0)[(\Gamma_A \cdot \Gamma_B
r^{-5}+
(3/2)m_Am_B\gamma_A^5 \gamma_B^5 r^{-3}] \eqno(6.9)$$
with $G^2 \equiv (1+c^2)G_AG_B$.  Thus, within a Dirac description of spin-1/2
particles,
 there is a term
 proportional to $m_Am_B r^{-3}$, which is of the type considered by Feynman in
studying the
 possibility that multi-neutrino exchange might lead to gravitation \cite{r26}.
However, it
comes
accompanied with a $\gamma^5$ factor for each particle.  Reduction of
$V_{2\nu}$ to
Schr\"{o}dinger-Pauli form
 then yields in the static limit only terms of order $r^{-5}$:
$$
	(V_{2\nu})_{red} = (G^2/4\pi^3  r^5)\{ 1+(\zeta_A\zeta_B/2)[5 {\bf \sigma}_A
\cdot {\bf \rh}
\, {\bf \sigma}_B \cdot {\bf \rh} -3 {\bf \sigma}_A \cdot {\bf \sigma}_B] \} \,
. \eqno(6.10)$$
The spin-independent part of (6.10) becomes $(2 \sin^2 \theta_W+1/2)
G_F^2/4\pi^3 r^5$
 in the standard model \cite{r10}.
 It is amusing to note that if one had two macroscopic bodies with an
appreciable fraction of
relativistic polarized electrons so that $\langle \gamma^5 \rangle \approx 1$
(hard to come by!),
 one could generate an
 effective interaction of the Feynman type.

\section{CONCLUDING REMARKS}

	We have seen that the concept of a potential in RQFT is rather subtle.  Once
one
 departs from the static approximation the demands of relativity inevitably
lead to velocity
dependence in the classical limit and, correspondingly, to energy dependence
and/or nonlocality
 of any potential designed to reproduce or facilitate the calculations of the
predictions of
 QFT.  Such dependence may be handled in a variety of ways.  I have sketched an
approach which
is based on the (perturbative) computation of the S-matrix.  If one confines
oneself to Feynman
 graphs which involve only the exchange of quanta between the interacting
systems, one may
define and compute systematically in perturbation theory a two-body potential
$V$ which is
 useful in the calculating the energies of bound states.  Such calculations
usually require a
 nonperturbative starting point, such as a Schr\"{o}dinger (or BS) type of
equation, even when
 the
 binding interaction is relatively weak, as in QED.  Nevertheless, the
potential (or kernel) to
 be used in such an equation may be computed perturbatively.

	 Some of the ambiguity in the potential is reduced by the requirement that it
reproduce
 the field-theoretic scattering amplitude not only in the physical region of
the scattering,
 but also outside this region, where the amplitude is defined by analytic
continuation.  The use of
 analyticity, when combined with generalized unitarity, also turns out to be a
powerful tool in
 computing the potential, especially for large separations $r$.  We have seen
how this method,
 originally developed for neutral spin-0 particles, can be used to obtain and
generalize old
results, such as that of Casimir and Polder on the retarded van der Waals
potential between
atoms, in a way which makes it clear why the result is universal, depending
only on general
features of field theory such as locality and gauge invariance.  Extension of
the method to the
case of a neutral particle and a charged particle leads to an effective
potential which has
proved to be useful in the analysis of Rydberg levels of helium atom.  We also
saw that the
extension of these methods to the case of two charged particles reveals a new
feature in the
concept of potential.  Different choices of the one-photon exchange potential
$V_{1\gamma}$
 may lead to
different results for the leading term in expansion in powers of $r^{-1}$ of
the two-photon
exchange
 potential $V_{2\gamma}$; this feature has a counterpart in classical
electrodynamics.
Recent extension
 of the method to spin-1/2 particles, either composite or elementary, has led
to formulas for
the $e^4$ corrections to the spin-orbit interaction and work on similar
corrections to the
spin-spin interaction is in progress.  When this is completed one will be able
to reanalyze the
 spin-dependent level structure of a number of two-body physical systems and
perhaps gain new
 insight into some aspects of bound-state QED, especially for states of large
orbital angular
momentum and relatively large separation between the constituents.

	On the conceptual side there are several issues which should be studied.  One
has to do
 with the incorporation of radiative corrections to bound states.  Because the
method is based
on the computation of on-shell matrix elements, one encounters IR divergences
if both
constituents are charged.  As long as one restricts attention to generalized
ladder graphs this
 problem is finessed by the computation of the potential, which is IR finite.
However, graphs
 with radiative corrections require the introduction of an IR cutoff and it
remains to be seen
 how this plays itself out in the computation of bound-state energies.  Even
apart from this,
it would be worth exploring the difference between the FGI and CGI potentials.
In second order
, the leading spin-independent parts of $V_{1\gamma}^{FGI}$ and
$V_{1\gamma}^{CGI}$  differ,
 by a term which has a
 classical character, while the spin-dependent parts are the same.  To what
extent do these
 features persist in higher orders?  An interesting theorem about QED may be
hidden here.
 The implications of a particular choice of $V_{1\gamma}$ and $V_{2\gamma}$ for
the accuracy
 achievable in the
study of many-body bound states also requires study, especially in connection
with a search for
 evidence of three-body forces.  Perhaps some progress on these issues will
have been made when
 next we meet in Minsk.

In conclusion, let me say that my visit to Belarus and the historic city of
Minsk is not without special emotions.  Many people I know can trace their
origins
to some of the small villages in Belarus, hundreds of which were systematically
destroyed by the Nazis in World War II.  Perhaps this meeting will play a
small part in the renaissance of Minsk as a center of learning and culture.

\begin{center}
Acknowledgments
\end{center}

	This work was supported in part by the U.S. National Science Foundation and by
a
 special grant from the American Physical Society. I wish to thank I.
Feranchuk,A. Gazizov, A. Gorbatsevich, O. Shadyro, Y. Shnir and L. Tomil'chik
for their help and splendid hospitality during my stay.


\begin{thebibliography}{99}
\bibitem{r1} C.G. Darwin, {\it Philos. Mag.} {\bf 39}, 537 (1920).
\bibitem{r2} G. Breit, {\it Phys. Rev.} {\bf 34}, 553 (1929).
\bibitem{r3} J.B. Gaunt, {\it Proc. R. Soc. London}, Ser. A {\bf 122}, 513
(1929).
\bibitem{r4} For a discussion of such ambiguities see J. Sucher, in  {\it
Proceedings of the
 Program on Relativistic, Quantum Electrodynamic and Weak Interaction Effects
in Atoms}, ITP,
 Santa Barbara (AIP Conf. Proc. No. 189, 1989), edited by W.R. Johnson, P.
Mohr, and J. Sucher (AIP, New York, 1989), p. 337.
\bibitem{r5} See, e.g. C. Itzykson and J.-B. Zuber, {\it Quantum Field Theory}
(McGraw Hill, New     York, 1980), p. 481.
\bibitem{r6} G. Feinberg and J. Sucher, {\it Phys. Rev.} A {\bf 2}, 2395
(1970).
\bibitem{r7} G. Feinberg and J. Sucher, {\it Phys. Rev.} A {\bf 27}, 1957
(1983).
\bibitem{r8} G. Feinberg and J. Sucher, {\it Phys. Rev.} D {\bf 38}, 3763
(1988).
\bibitem{r9} R. Cutkosky, {\it J. Math. Phys.} {\bf 1}, 429 (1960).
\bibitem{r10} G. Feinberg, J. Sucher, and C.K. Au, {\it Phys. Rep.} {\bf 180},
859 (1989).
\bibitem{r11} H.B.G. Casimir and D. Polder, {\it Phys. Rev.} {\bf 73}, 360
(1948).
\bibitem{r12} M. O'Carroll and J. Sucher, {\it Phys. Rev.} {\bf  182}, 85
(1969).
\bibitem{r13} S. C. Wang, {\it Phys. Z.} {\bf 28}, 663 (1927);  F. London, {\it
Z. Phys. Chem.}
{\bf 11}, 222 (1930).
\bibitem{r14} J. Bernabeu and R. Tarrach, {\it Ann. Phys.} (N.Y.) 102, 323
(1976).
\bibitem{r15} E. Kelsey and L. Spruch, {\it Phys. Rev.} {\bf 18}, 15 (1978);
18, 1055 (1978).
\bibitem{r16} S. Lundeen, in {\it Long-Range Casimir Forces: Theory and Recent
Experiments in
       Atomic Systems}, edited by F.S. Levin and D.A. Micha (Plenum, New York,
1993).
\bibitem{r17} C.-K. Au, G. Feinberg, J. Sucher, {\it Phys. Rev. Lett.} {\bf
173}, 355 (1987);
    G. Feinberg, J. Sucher, and C.K. Au, {\it Ann. Phys.} {\bf 173}, 355
(1987).
\bibitem{r18} J. Sucher, {\it Phys. Rev.} D {\bf 49}, 4284 (1994).
\bibitem{r19} J. Sucher, {\it Comm. At. Mol. Phys.} {\bf 30}, 129 (1994).
\bibitem{r20} G. Brown and D.G Ravenhall, {\it Proc. R. Soc. London} Ser. A
{\bf 208}, 552 (1951).
\bibitem{r21} J. Sucher, {\it Phys. Rev.} A {\bf 25}, 348 (1980); {\it Phys.
Rev. Lett.} {\bf 55}
, 1023 (1985).
\bibitem{r22} G. Feinberg and J. Sucher. {\it Phys. Rev.} D  {\bf 45}, 2493
(1992).
\bibitem{r23} G. Gilbert and J. Sucher, work in progress.
\bibitem{r24} G. Feinberg and J. Sucher, {\it Phys.Rev.} {\bf 166}, 1638
(1965).
\bibitem{r25} J. Sucher, in Proceedings of the XIVth Moriond Workshop, {\it
Particle Astrophysics,
      Atomic Physics and Gravitation}, Villars-sur-Ollon, January 1994 (to
appear).
\bibitem{r26} R.P. Feynman, {\it Lectures on Gravitation} , 1962-63; notes by
F.B. Morinigo and
    W.W. Wagner (California Institute of Technology, 1971).
\end{thebibliography}
\end{document}